\pdfoutput=1
\documentclass[a4paper,12pt]{spieman}  
\usepackage{amsmath,amsfonts,amssymb}
\usepackage{graphicx}
\usepackage{setspace}
\usepackage{tocloft}

\usepackage{siunitx}
\usepackage[usenames,dvipsnames]{xcolor}
\usepackage{wasysym}
\usepackage{hyperref}
\usepackage{lettrine}
\usepackage{fix-cm} 
\usepackage{bm}

\newcommand{\added}[1]{{\color[RGB]{0,0,0}{#1}}}

\begin{document}
\title{Wideband on-chip terahertz spectrometer based on a superconducting filterbank}

\author[a,b,*]{Akira~Endo}
\author[c,a]{Kenichi~Karatsu}
\author[c,a]{Alejandro~Pascual~Laguna}
\author[b]{Behnam~Mirzaei}
\author[c]{Robert~Huiting}
\author[a,b]{David~J.~Thoen}
\author[c]{Vignesh~Murugesan}
\author[d]{Stephen~J.~C.~Yates}
\author[c]{Juan~Bueno}
\author[a]{Nuri~van~Marrewijk}
\author[a]{Sjoerd~Bosma}
\author[a]{Ozan~Yurduseven}
\author[a]{Nuria~Llombart}
\author[e]{Junya~Suzuki}
\author[f]{Masato~Naruse}
\author[c]{Pieter~J.~de~Visser}
\author[g]{Paul~P.~van~der~Werf}
\author[h]{Teun~M.~Klapwijk }
\author[c, a]{Jochem~J.~A.~Baselmans}
\affil[a]{Faculty of Electrical Engineering, Mathematics and Computer Science, Delft University of Technology, Mekelweg 4, 2628 CD Delft, the Netherlands.}
\affil[b]{Kavli Institute of NanoScience, Faculty of Applied Sciences, Delft University of Technology, Lorentzweg 1, 2628 CJ Delft, The Netherlands.}
\affil[c]{SRON---Netherlands Institute for Space Research, Sorbonnelaan 2, 3584 CA Utrecht, The Netherlands.}
\affil[d]{SRON---Netherlands Institute for Space Research, Landleven 12, 9747 AD Groningen, The Netherlands.}
\affil[e]{High Energy Accelerator Research Organization (KEK), 1-1 Oho, Tsukuba, Ibaraki, 305-0801, Japan.}
\affil[f]{Graduate School of Science and Engineering, Saitama University, 255, Shimo-okubo, Sakura, Saitama 338-8570, Japan.}
\affil[g]{Leiden Observatory, Leiden University, PO Box 9513, NL-2300 RA Leiden, The Netherlands.}
\affil[h]{Physics Department, Moscow State Pedagogical University, 119991 Moscow, Russia.}

\renewcommand{\cftdotsep}{\cftnodots}
\cftpagenumbersoff{figure}
\cftpagenumbersoff{table} 
\maketitle

\begin{abstract}
Terahertz spectrometers with a wide instantaneous frequency coverage for passive remote sensing are enormously attractive for many terahertz applications, such as astronomy, atmospheric science and security. Here we demonstrate a wide-band terahertz spectrometer based on a single superconducting chip. The chip consists of an antenna coupled to a transmission line filterbank, with a microwave kinetic inductance detector behind each filter. Using frequency division multiplexing, all detectors are read-out simultaneously creating a wide-band spectrometer with an instantaneous bandwidth of 45~GHz centered around 350~GHz. The spectrometer has a spectral resolution of $F/\Delta F=380$ and reaches photon-noise limited sensitivity. We discuss the chip design and fabrication, as well as the system integration and testing. We confirm full system operation by the detection of an emission line spectrum of methanol gas. The proposed concept allows for spectroscopic radiation detection over large bandwidths and resolutions up to $F/\Delta F\sim1000$, all using a chip area of a few $\mathrm{cm^2}$. This will allow the construction of medium resolution imaging spectrometers with unprecedented speed and sensitivity.
\end{abstract}


{\noindent\footnotesize\textbf{*}Akira~Endo, \linkable{a.endo@tudelft.nl}}

\begin{spacing}{1}   

\begin{figure*}
 \includegraphics[width=\textwidth]{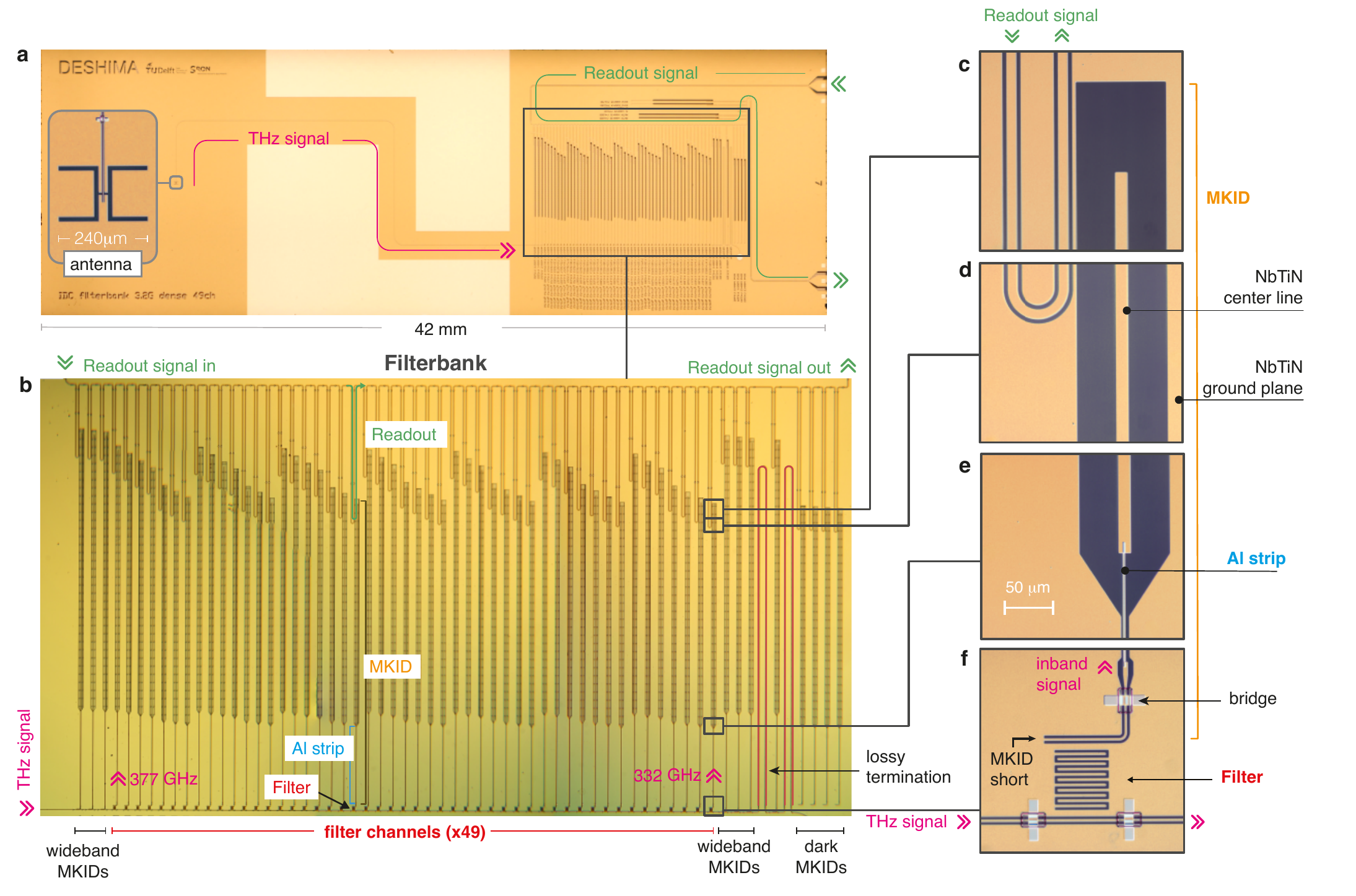}%
 \caption{\label{fig:chip} 
 \textbf{Filterbank spectrometer chip.}  
  \textbf{a}, Micrograph of the entire spectrometer chip, which has dimensions of 42~mm~$\times$~14~mm. \added{The features above the filterbank include two 'dark MKIDs,' as well as MKIDs with center lines made fully of either NbTiN or Al for health-check purposes. The features below the filterbank are text labels patterned in the NbTiN ground plane.}
  \textbf{b}, Close-up view of the filterbank circuit. The spectral channels are placed in decreasing order of frequency, from maximum (377~GHz) to minimum (332~GHz). At the top left and right corners of the filterbank are the input and output ports of the microwave readout signals, respectively. 
  \added{By design, the readout resonance frequency of the 49 filter MKIDs are first distributed equally between 5.6 and 6.4 GHz, with a step of $\Delta F = 16.7$  MHz. The MKIDs are further grouped into seven groups of seven MKIDs: from left to right, the MKID in group $g$ (=0,1,...,6) of index $i$ (=0,1,...,6) has a resonance frequency of $f_{g,i} = 5.6+ (g+ 7i)\Delta F$~GHz. Finally, the frequencies below and above 6.0 GHz are shifted by $-$0.1 GHz and $+$0.1 GHz, respectively, to allow the placement of the local oscillator tone at 6.0 GHz. As a result, the resonance frequencies are in two ranges of 5.5--5.9 and 6.1–-6.5 GHz.}
  \added{The MKID between the 2 segments of lossy termination, which are highlighted in red, is a 'wideband MKID' to help health-check that the signal is properly absorbed by the lossy termination.} 
  \textbf{c--f,} Further close-up view of one of the spectral channels of the filterbank. The filter (\textbf{f}) is an interdigitated pattern etched in NbTiN, acting as a resonant bandpass filter. On both sides of the filter, the ground planes of the terahertz line are shorted with aluminium bridges over a block of UV-patterned polyimide. A bridge is also placed on the NbTiN CPW coupler of the MKID. Every MKID (\textbf{c--f}) consists of an short-ended NbTiN CPW that couples to the filter (\textbf{f}), a narrow CPW with an aluminium center line(\textbf{e, f}), and an open-ended wide NbTiN CPW (\textbf{c--e}). The MKID is coupled to the microwave readout line near the open end (\textbf{c, d}).
  }%
\end{figure*}

\section{Introduction}

Cool gas is the most abundant phase of matter in the interstellar medium of galaxies, and in the Earth's atmosphere. Hence, precise knowledge of the spatial variation and dynamics of the physical and chemical properties of cool gas is of crucial importance for astrophysics\cite{Kulesa:kl}, as well as weather forecasting and global climate modeling\cite{2017SPIE10210E..10H, Prigent:2006hl}.
These properties can be diagnosed using passive terahertz remote sensing, because many molecules and atoms in cool gas make energy level transitions that leave distinct fingerprints in the terahertz emission spectrum\cite{Walker:2015tq}. However, most of the currently used \textit{coherent} receivers\cite{Graf:2015eoa} can observe only $\sim$1--10~GHz of bandwidth at a time (exceptionally up to $\sim$35~GHz\cite{Erickson:2007ub, Primiani:2016ef}). This makes it very time consuming to cover a significant fraction of the 0.1--1~THz band (referred to as `terahertz radiation' in the remainder of this article), where the atmosphere is partially transparent. Indeed, it is often essential to probe multiple energy transitions of multiple chemical species to make a meaningful diagnosis, calling for much wider simultaneous bandwidths\cite{2017SPIE10210E..10H,2011ITTST...1..241S}. 

The on-chip filterbank spectrometer\cite{2012JLTP..167..341E,Endo:2013ky,Wheeler:2018cg,Endo:2015hj} is a recently proposed concept that aims to enable wideband terahertz remote-sensing spectroscopy with a threefold advance. The first step is to use an array of \textit{incoherent} detectors\cite{2011ITTST...1..241S} to measure the signal, after dispersion, in one detector for each frequency bin. This decouples the observation bandwidth from the detector readout bandwidth, allowing very efficient back-ends for medium resolution spectroscopy. The second step is to use superconducting millimeter wave electronics to: i) couple the signal to the chip, ii) \added{disperse} it and iii) \added{measure} it. This integrates the functionality of a classical dispersion spectrometer\cite{2011ITTST...1..241S} with a typical size of $\le$1 m onto a chip of a few $\mathrm{cm^2}$, which makes the spectrometer much more scalable towards larger bandwidths, longer wavelengths, and a larger number of spectrometers to form a spectral-imaging array. The emergence of such a passive spectroscopic imager could also reform the landscape of terahertz applications in our daily lives, because it combines the advantages of a broad input frequency band, ideally suited for pressure broadened lines, fast imaging without scanning, real-time material diagnosis even through clothing, and not irradiating the subject\cite{Kanda:2017js, Tonouchi:2007cs, Dhillon:2017dza,  Appleby:2007gs}. The last advance is the fact that incoherent detectors allow for extremely sensitive \added{spectrometers}, especially for low radiation environments, because they are not subject to quantum noise\cite{2011ITTST...1..241S}.

Here we demonstrate the design, performance and operation of \added{a} terahertz on-chip filterbank spectrometer system, consisting of a spectrometer chip, cryogenic system and readout electronics. The spectrometer chip, as shown in Fig. \ref{fig:chip}, covers a frequency band of 332--377~GHz with 49 spectral channels with a constant spectral resolving power of $F/\Delta F = 380$. It is based on incoherent NbTiN-Al hybrid microwave kinetic inductance detector (MKID) technology\cite{2016JLTP..184..412E} that can be scaled to \added{near the NbTiN gap frequency of $\sim$1.1 THz.} The sensitivity we demonstrate here is set by the photon noise inherent to the incoming optical signal, and by the instrument radiation coupling efficiency. Using large bandwidth antennas such as the leaky wave antenna \cite{2017ApPhL.110w3503B} and the sinuous antenna\cite{OBrient:2013hc}, input bandwidths in excess of one octave are possible (e.g., 300--900~GHz), allowing truly wideband spectroscopy with up to $\sim$1000 channels per spectrometer.

\section{Filterbank spectrometer chip}

 The chip is fabricated from a 100~nm-thick NbTiN film, which is deposited on the c-plane sapphire substrate using reactive magnetron sputtering \cite{Thoen:2017bc}. \added{This film has a critical temperature $T_\mathrm{c} = 15\ \mathrm{K}$, which implies that the material has low losses up to near the gap frequency $F_\mathrm{gap}\sim 3.52 k_\mathrm{b}T_\mathrm{c}/h=1.1\ \mathrm{THz}$.} ($k_\mathrm{b}$ is the Boltzmann constant, and $h$ is the Planck constant.) The chip couples to linearly polarized radiation using a double slot antenna, patterned using UV contact lithography and plasma etching in a $\mathrm{SF_6}+\mathrm{O_2}$ plasma, followed by an in-situ $\mathrm{O_2}$ cleaning\cite{Ferrari:bk}. The antenna is placed in the focus of a $\diameter$8 mm extended-hemispherical Si lens\cite{2017A&A...601A..89B}, which is anti-reflection coated with a 130 $\mathrm{\mu m}$-thick layer of Parylene-C \cite{2000stt..conf..407J}. The received signal is guided to the filterbank through the terahertz line: a coplanar waveguide (CPW) patterned in the NbTiN ground plane. The filterbank sorts the wideband signal into 49 sub-bands by means of narrow band pass filters. These filters are coplanar, interdigitated resonators, as shown in Fig.~\ref{fig:chip}f, which are coupled to the terahertz line on one side, and to a NbTiN-Al hybrid  MKID\cite{Janssen:2014fp} on the other side. At the resonant frequency of the filter, maximum signal power is transferred from the terahertz line to the MKID. Each MKID is a CPW quarter wavelength resonator, with a resonant frequency in the order of 5~GHz. Its open end is coupled to the readout line and its shorted end is coupled to the terahertz filter. Near the terahertz filter the MKID CPW has a central line made from 40 nm-thick aluminium with a resistivity of 0.8 $\mathrm{\mu \Omega\ cm}$ and $T_\mathrm{c} = 1.25\ \mathrm{K}$.  For the signal frequencies of 332--377~GHz the aluminium acts as a radiation absorber, because $F_\mathrm{gap, Al}=90\ \mathrm{GHz}$ is smaller than the signal frequency band of the spectrometer. The radiation power absorbed creates a proportional shift in the resonant frequency of the MKID, which is read out as a change in the transmission phase of a microwave tone in the readout line that connects to all MKIDs. The ground planes of the readout line, and the terahertz line, are balanced using aluminium bridges with dielectric bricks fabricated from spin-coated polyimide LTC9505 from Fujifilm\cite{Ferrari:bk}.As a reference for the terahertz signal power at the input of the filterbank, three `wideband MKIDs' are placed before the filterbank; these are MKIDs that are weakly coupled to the terahertz line directly without a filter, with a nearly constant power coupling of $-27$~dB over the relevant frequency range. Similarly, another three wideband MKIDs are placed after the filterbank for a reference of the power that runs through the filterbank without being drawn out by the filter channels. Furthermore, there are four `dark MKIDs' that are placed away from the terahertz line as a reference for signal power coupled to the MKIDs by way of stray light or surface waves\cite{Yates2017}. The chip is equipped with a mesh of 40~nm-thick $\beta$-phase Ta on its backside to reduce the propagation of stray radiation inside the chip. The terahertz signal line is terminated by a CPW with an aluminium center line, which absorbs the remaining terahertz signal after the filterbank, to prevent reflection of power that is not drawn out by the filter channels. 

 \begin{figure*}
 \includegraphics[width=\textwidth]{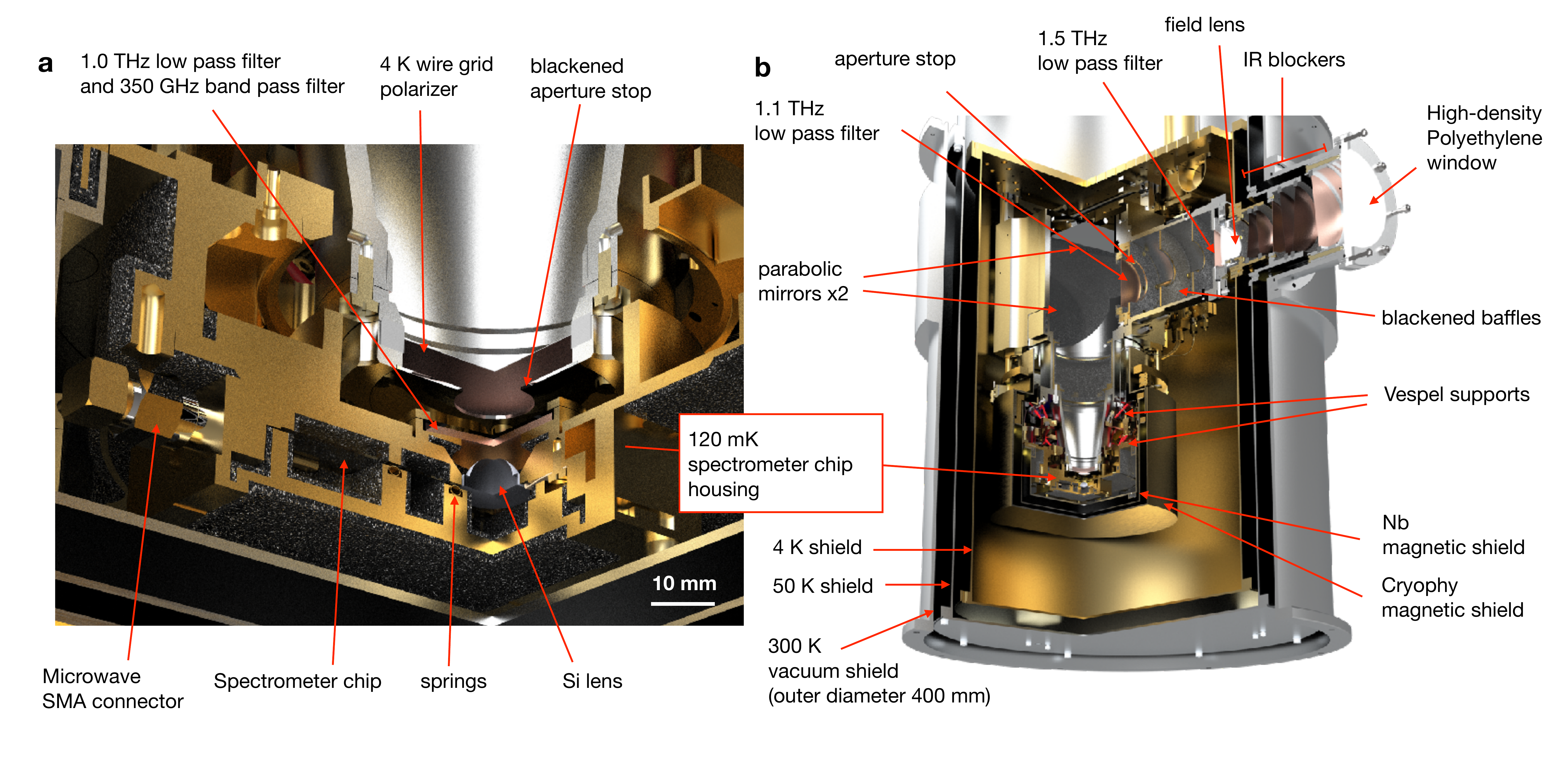} %
 \caption{\label{fig:cryostat} 
 \textbf{Cryo-optical setup of the spectrometer.}  
  \textbf{a}, Quarter-cutout, cross sectional view of a computer-aided drawing of the spectrometer chip mounted in a housing made of gold-plated copper. The chip is laid horizontally, with the side of the spectrometer circuit facing down. The Si lens looks into two quasioptical filters, one of which is a 1.0 THz low pass filter and the other is a 350~GHz band pass filter.
  The interior of the chip holder is coated with carbon loaded epoxy loaded with SiC grains\cite{Baselmans2012UltraLowBG}. The readout signal is sent in and out from the SMA connector (only one of the two is visible due to the cross-section).
  \textbf{b}, Quarter-cutout, cross-sectional view of  the cryo-optical structure. The detector housing is placed at 120 mK, near the bottom of a double-layer cup structure made of Cryophy and Nb. The 4 K optics tube holds a wire grid, a pair of parabolic mirrors, two low pass filters, a field lens, and a pair of infrared blockers. There are also infrared blockers on the 50 K shield and 300 K shield.
 }%
 \end{figure*}

\section{Spectrometer System}

The filterbank spectrometer chip is integrated into a system, which includes the detector readout using microwave frequency division multiplexing (FDM) for simultaneously reading out all detectors, optics to couple the terahertz signal to the chip, and a refrigerator for cooling them. The spectrometer chip is mounted in a housing as displayed in Fig.~\ref{fig:cryostat}a. The optical entrance of the chip housing is closed off above the lens with a 1~THz low-pass filter, and a 350~GHz bandpass filter with a transmission as shown later in Fig.~\ref{fig:grid}e. The chip housing is mounted inside a light-tight chamber equipped with coaxial cable filter feedthroughs\cite{Baselmans2012UltraLowBG}. The chamber is cooled via copper thermal straps to 120~mK using a two-stage adiabatic demagnetisation refrigerator (ADR), connected to the 4 K stage provided by a two-stage pulse tube cooler. The chamber itself is isolated from the 4~K environment by a thermal-mechanical structure based on Vespel tubes. Both the chamber and chip housing operated at 120~mK are coated on the inside by a radiation absorber that consists of 3\% by weight carbon powder mixed with Epotek~92 epoxy, in which we embed 1~mm or 0.5~mm SiC grains as diffusive scattering elements. This recipe is adapted from Ref.\cite{2001stt..conf..400K} to yield better absorption at long wavelengths and better adhesion on large surfaces. A microwave analog/digital readout system\cite{Rantwijk2016} and a 4~K low noise  amplifier are used to read out the phase response of all of the MKIDs simultaneously. This is achieved by creating a time dependent signal that is the reverse Fourier transform of a set of single frequency tones, one for each MKID, with the addition of a few `blind' tones used for system calibration and monitoring. These tones are sent to the chip, where radiation absorbed in an MKID modifies its associated single frequency tone. After passing through the chip the signal is amplified at 4~K using the low noise amplifier with a noise temperature of 5~K, after which it is further amplified and analyzed in the readout system. Here the change in complex transmission at each tone is obtained using an on-board fast Fourier transform (FFT) engine, and it is converted to a resonance frequency response in the post analysis to increase the linearity\cite{Bisigello:2016ff}. Radiation is coupled to the detector chip using a 4~K light-tight box, equipped with an optical relay of two parabolic mirrors, baffling structures coated with the same radiation absorber as the 120~mK box and sample holder, infrared filters and a polarizing filter mounted in co-polarization with the antenna, as shown in Fig.~\ref{fig:cryostat}b. The relay creates a pupil where the beam is tapered at the $-$10~dB level, limiting large angular stray radiation. 
\added{Even in a single beam system the pupil has a specific meaning in optics, being the conjugate plane of the focal plane which limits the angular extent of the beam. Complete baffling to limit the optical throughput incident on the chip in both area extent and angle therefore requires limiting apertures (stops) in both conjugate planes (focal and pupil planes). Here aperture baffles at the window, on the array and the lens itself act as an effective focal plane aperture stop. While at the pupil there is an aperture limiting the angle to a f/D ratio of 3.9 (7.7 deg.)}
This entire optical chain is cooled to 4~K. In addition, the chip is surrounded by two layers of magnetic shielding (Nb and Cryophy\cite{Aperam:EUrtQmw7}), also cooled to 4~K. The combined shields attenuate the magnetic field perpendicular to the chip by a factor $\sim$700.

\begin{figure*}
 \includegraphics[width=0.8\textwidth]{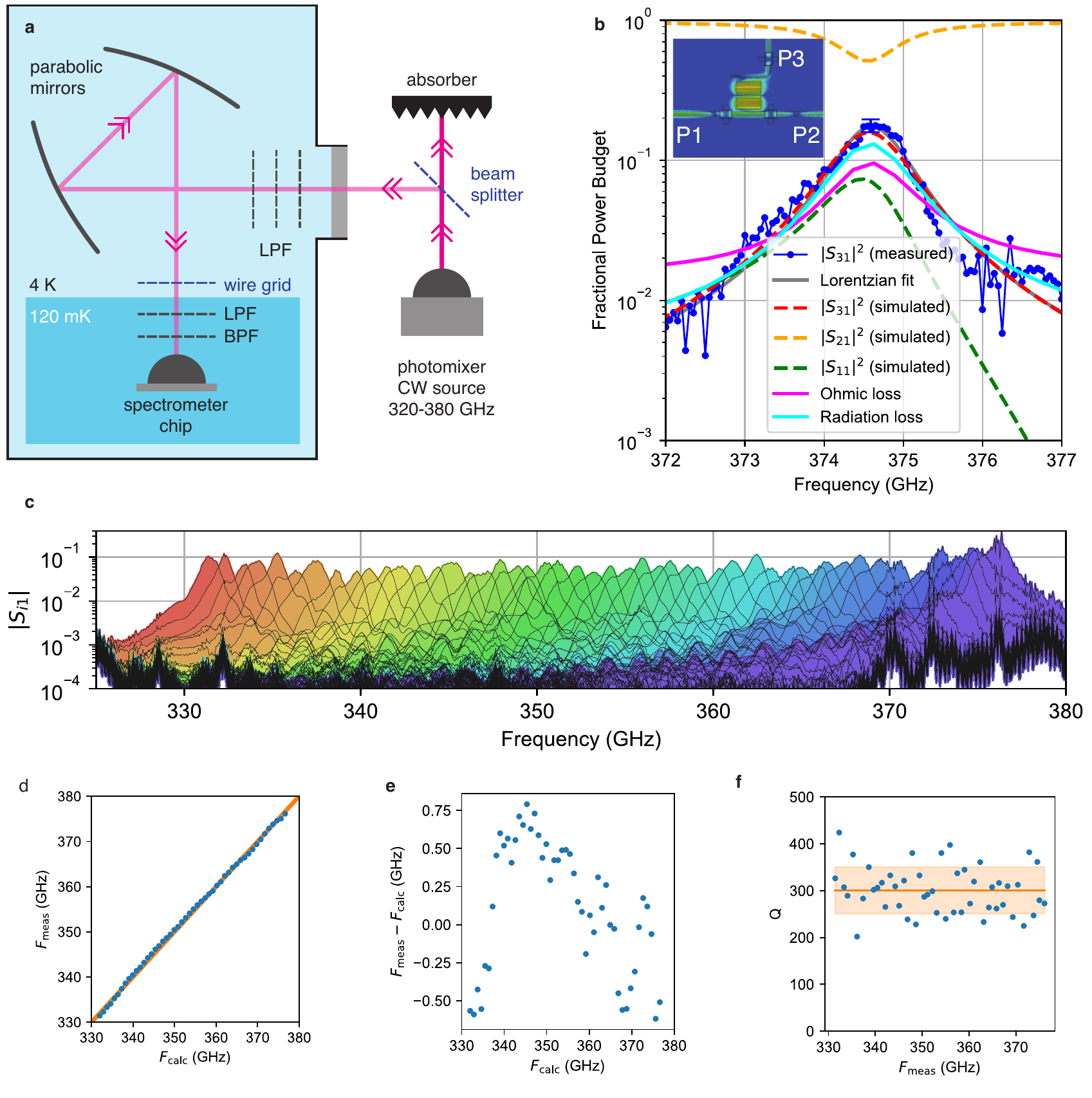} 
 \caption{\label{fig:toptica} 
 \textbf{Terahertz frequency response of the spectrometer chip.}  
  \textbf{a,} The experimental configuration for measuring the frequency response. The terahertz CW signal beam from the photomixer source is reflected off a beam splitter made of a sheet of Mylar. The reflected signal enters the cryostat into the 4~K optics. The beam passes through a stack of low pass filters (LPF), two parabolic mirrors, a wire grid, and finally the 120 mK low pass filter and bandpass filter (BPF) before it reaches the spectrometer chip.
    \textbf{b,} The measured and simulated frequency-dependent 3-port scattering parameters. 
    \added{At the peak of the measured $|S_{31}|^2$ we indicate an error bar derived from the error in the optical efficiency measurement shown in Fig.~\ref{fig:grid}e.}
    \added{The simulation is for one filter in isolation, with the 3 ports terminated with a matched load.} The inset shows the simulated electric field distribution $|{\bm E}|^2$ around the filter, together with the port definitions: 
    Port 1 is the input from the terahertz through line, port 2 is the output towards the subsequent filters, and port 3 is the output towards the MKID. The plot also includes the simulated fractional power absorbed by ohmic loss in the aluminium bridges, and radiation loss into the substrate. The measured data are fitted with a Lorentzian curve, yielding a $Q$-factor of 361 for this filter. 
	\textbf{c,} Measured scattering parameters $|S_{i1}|^2$ of all 49 channels ($i=3,4,...,52$), from the filterbank entrance (port 1) to the MKID.
    \textbf{d,} The measured peak frequency $F_\mathrm{meas}$, determined from individual Lorentzian curve fitting, compared to an ideal geometric sequence of $F_\mathrm{calc} = 332\times (1+\Delta F/F)^i$, where the frequency spacing is $F/\Delta F=380$ and $i$ is the channel index. The solid line indicates $F_\mathrm{meas}=F_\mathrm{calc}$.
    \textbf{e,} Residual of $F_\mathrm{meas}-F_\mathrm{calc}$, plotted as a function of $F_\mathrm{calc}$.
     \textbf{f,} Filter $Q$ determined from a Lorentzian fitting to the peaks, plotted against $F_\mathrm{meas}$. The solid line and the shaded area indicate the mean and standard deviation of $Q$, respectively.
 }%
 \end{figure*}

\section{Spectral response}

The frequency response of the spectrometer is measured with a photomixing terahertz continuous wave (CW) source (Toptica Terabeam 1550) in a setup as shown in Fig.~\ref{fig:toptica}a. We couple the beam from the CW source to the cryostat by reflecting it from a 15 $\mathrm{\mu m}$-thick sheet of Mylar to attenuate the signal power by $-24$~dB to $-25$~dB. We then measure the response of each MKID while sweeping the frequency of the CW source from 320~GHz to 380~GHz. In all results we apply a frequency shift of $-0.56$~GHz to the CW source frequency as a result of a frequency calibration of the CW source using a methanol emission line as will be discussed in Sec. \ref{sec:gascel}. 
Fig.~\ref{fig:toptica}b shows the response of one filter channel MKID near the filterbank entrance, which has less influence from neighboring channels and transmission losses compared to filters further downstream. 
\added{The calculation of the response takes the following steps:
i) divide the frequency-dependent response of the filter-MKID with the response of a wideband MKID at the entrance of the filterbank (see Fig.\ref{fig:chip}); 
ii) normalize the peak of the response to unity to calculate the instrument optical efficiency ($\eta_\mathrm{inst}$), see Fig.~\ref{fig:grid}e. 
iii) return to the frequency-dependent response and adjust it so that the peak height is equal to $\eta_\mathrm{inst}$ divided by the losses between the entrance of the filterbank to the cryostat window.
In this way the only result from the photomixing-source scan is the relative frequency response, and the peak height of $17.7\% \pm 1.8\%$ correctly represents the magnitude of the scattering parameter $|S_{31}|^2$ according to the definition in the inset of Fig.~\ref{fig:toptica}b. }
A Lorentzian fit to the measured transmission peak yields a quality factor of $Q=361$.
\added{From the measured $|S_{31}|^2$ peak and $Q$ we can infer an internal $Q_\mathrm{i} =891$ and a coupling $Q_\mathrm{c}$ of $1.21\times 10^3$, assuming the same coupling strength to the feed line and to the MKID\cite{Endo:2013ky}.}
\added{The simulation using CST Microwave Studio shows that} the filter has a measured transmission to the MKID of $|S_{31}|^2=16\%$ on resonance, which is consistent with the \added{measured value}. \added{The peak transmission is} smaller than the theoretical maximum transmission of 50\% for a single bandpass filter\cite{Shirokoff:2012fx}. According to the power budget around the resonance frequency simulated with CST Microwave Studio\cite{CSTMicrowaveStudio} shown in Fig.~\ref{fig:toptica}b, this can be understood as a result of the following contributions: 
$|S_{21}|^2(\mathrm{simulated})$~=~51\% of the power passing through the terahertz signal line; 
$|S_{11}|^2(\mathrm{simulated})$~=~7\% being reflected back towards the input of the filterbank; 
13\% being lost as radiation from the filter into the substrate; 
and 10\% being absorbed by the aluminium bridges; 
leaving $|S_{31}|^2(\mathrm{simulated})=16\%$ to couple to the MKID. 
\added{Note that the simulated coupling numbers do not add up to 100\%, but to 97\%, due to the total power budget from the CST simulation. This uncertainty is much smaller than the error in the absolute efficiency measurement.}

The measured transmission from the filterbank entrance to the MKIDs for all 49 channels is plotted in Fig.~\ref{fig:toptica}c. With $F/\Delta F=380$ being the frequency spacing between adjacent channels, the peak frequencies of the filters follow well a geometric sequence of $F_{i}=332\times(1+\Delta F/F)^i$~GHz, as presented in Fig.~\ref{fig:toptica}d. The convex trend seen in the small residual as presented in Fig.~\ref{fig:toptica}e is likely due to a gradient in sheet inductance of the NbTiN film along the filterbank \cite{Thoen:2017bc}, and a scatter on the order of 0.1\% of the resonance frequency as seen here is typical for \added{the readout frequency of} MKID arrays\cite{2017A&A...601A..89B,2018arXiv180304275M}. In Fig.~\ref{fig:toptica}f we present the quality factors of all channels, showing an average $ \overline{Q} =300\pm50$. The frequency spacing of $F/\Delta F = 380$ is an intentional oversampling compared to the mean filter $Q$ of 300, to conservatively avoid gaps in the spectral coverage (this resolution is well suited for the detection of redshifted emission lines from terahertz-bright galaxies\cite{Carilli2006}). As a consequence of the power being partially shared among neighboring channels, the oversampling causes the peak transmission of many channels to be slightly lower (mean of 8\%) than the highest-frequency channels close to the entrance of the filterbank, such as the one shown in Fig.~\ref{fig:toptica}b. Collectively, the sufficiently small scatter in both filter frequency and $Q$ \added{with a channel yield of 100\%} makes it possible to cover the entire 332--377~GHz band without significant gaps, as demonstrated in Fig.~\ref{fig:toptica}c.

\section{System sensitivity}

\begin{figure*}
 \includegraphics[width=\textwidth]{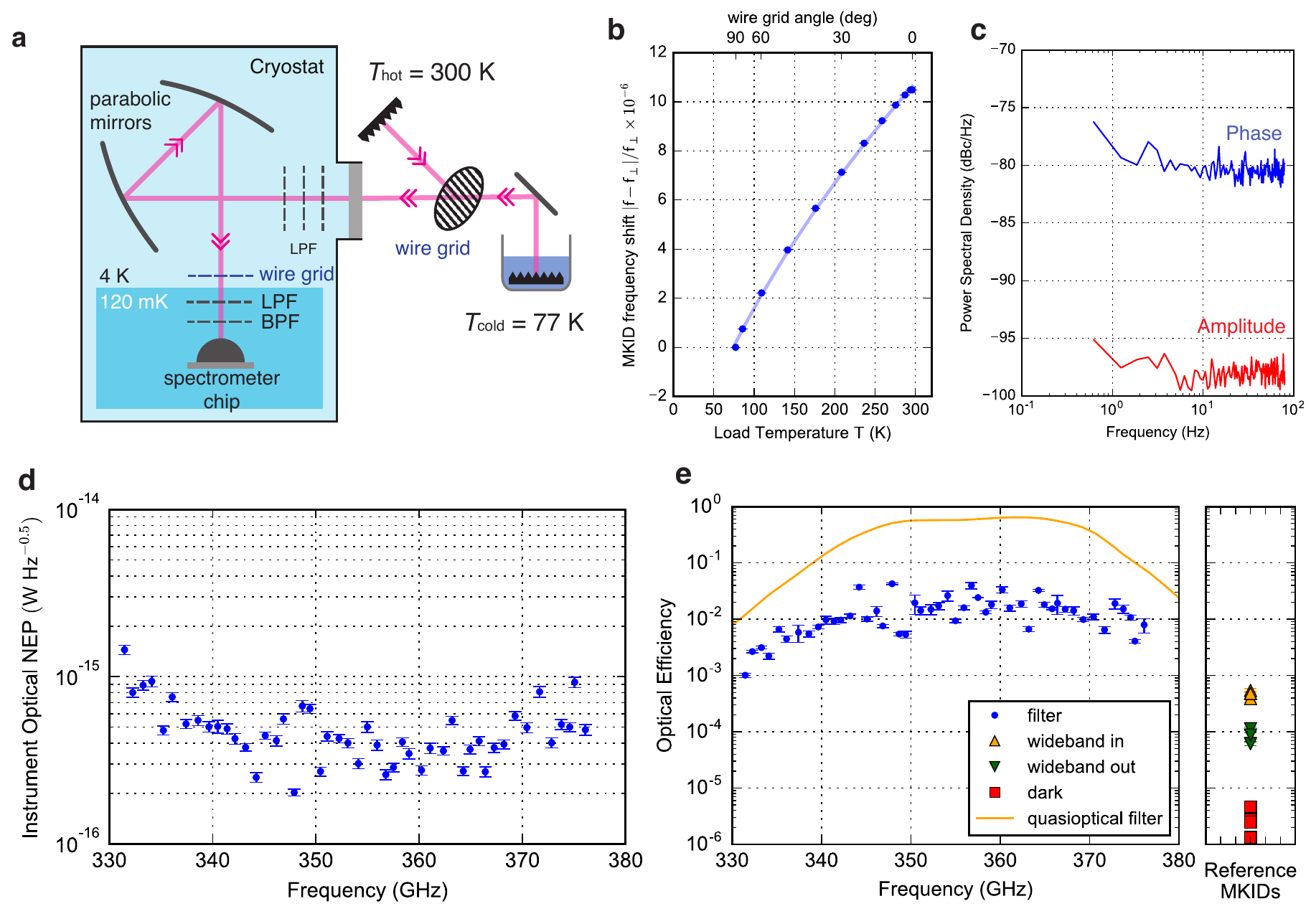} %
 \caption{\label{fig:grid} 
 \textbf{Instrument sensitivity of the filterbank spectrometer.}  
  \textbf{a}, Optical setup for characterizing the instrument sensitivity. The cryogenic wire grid transmits the linear polarization to which the antenna on the spectrometer chip is sensitive. The room temperature wire grid is rotated so that the spectrometer sees a mixture of radiation from the $T_\mathrm{hot}=300\ \mathrm{K}$ black body, and the $T_\mathrm{cold}=77\ \mathrm{K}$ black body immersed in liquid $\mathrm{N_2}$. 
  \textbf{b}, Response in relative MKID resonance frequency shift of one of the filterbank channels, as a function of the angle of the room temperature wire grid, and the resulting effective load temperature looking into the grid. 
  The resonance frequency shift is measured relative to the resonance frequency ($f_\perp$) at which the wires are perpendicular to the polarization the spectrometer is sensitive to, and the spectrometer sees only the cold black body. 
  The curve is a \added{square-root-law} fit to the data points. 
  \textbf{c}, Measured phase and amplitude noise power spectral density (PSD) of a representative filterbank channel MKID, measured at minimum load temperature of $T_\mathrm{load}=T_\mathrm{cold}=77\ \mathrm{K}$. 
  Note that the spectra are taken from the signal relative to the circle traced by the frequency sweep around the resonator; see \cite{Gao:2008ks} for details.
   \textbf{d}, Optical Noise Equivalent Power for all filterbank channels as a function of filter peak frequency. The error bars (1 standard deviation) are combined statistical uncertainties from the responsivity and noise.
  \textbf{e}, Optical efficiency of all filter channels (left) and reference MKIDs \added{(right)} as a function of the transmission peak frequency. The error bars (1 standard deviation) are combined statistical uncertainties from the responsivity and noise. The left panel also shows the transmission of the quasioptical filter stack as a function of frequency (solid curve).
 }%
\end{figure*}

Because the MKID detectors used in this study are photon-noise limited in the relevant range of terahertz loading power \cite{Janssen:2014fp}, we can measure the sensitivity of the spectrometer by using the stream of photons as an absolute calibration source. A pair of black body radiators at room temperature $T_\mathrm{hot}=300\ \mathrm{K}$ and liquid $\mathrm{N}_2$ temperature  $T_\mathrm{cold}=77\ \mathrm{K}$ are placed at reflection and transmission positions of a wire grid seen from the spectrometer window, respectively, as shown in Fig.~\ref{fig:grid}a. Looking into the wire grid, the spectrometer sees an effective load brightness temperature of $T_\mathrm{load} = rT_\mathrm{hot} + (1-r) T_\mathrm{cold}$, where $r$ is the reflectance in co-polarization to the cryogenic wire grid and on-chip antenna. 

Fig.~\ref{fig:grid}b shows the relative frequency response $x\equiv (f_\perp - f) /f_\perp$ ($f_\perp$ is the MKID resonant frequency for $T_\mathrm{load} = T_\mathrm{cold}$) of a representative filter-channel MKID, as a function of $T_\mathrm{load}$. From a square-root-law fit \added{(expected if the responsivity is scaling as $P_\mathrm{abs}^{0.5}$ where $P_\mathrm{abs}$ is the absorbed power\cite{2014NatCo...5E3130D}, and the total optical loading is dominated by the variable thermal load outside of the cryostat)}, we determine the temperature response  $x(T_\mathrm{load})$. To be able to measure the spectrometer sensitivity expressed in noise equivalent power (NEP), we convert the load temperature to radiation power $P_\mathrm{rad}(T_\mathrm{load})$, which is defined as the single-mode, single polarization radiation power outside of the cryostat window that can couple to a single filter channel:
\begin{equation}\label{eq:P}
		P_\mathrm{rad}(T_\mathrm{load}) = \frac{1}{2}\int^\infty_0 \frac{c^2 t(F) B(F,T_\mathrm{load})}{F^2} dF.
\end{equation}
Here,  $t(F)$ is the transmission of the filter channel, as shown in Fig.~\ref{fig:toptica}c, with the peak transmission normalized to unity. Note that we have limited the integral bounds of Eq.\ref{eq:P} to the bandwidth of the quasioptical filter to avoid integrating noise as signal. Furthermore, $c$ is the speed of light in vacuum, and $B(F, T_\mathrm{load})$ is the Planck brilliance as a function of $T_\mathrm{load}$ and frequency $F$. The NEP can now be determined experimentally by combining the responsivity $\frac{dx}{dP_\mathrm{rad}}$ with MKID photon noise level $S_x$ using 
\begin{equation}\label{eq:NEP}
\begin{split}
	NEP_\mathrm{exp}(T_\mathrm{load}) &= \sqrt{S_x}\biggl( \frac{dx}{dP_\mathrm{rad}} \biggr)^{-1}\\ 
	&= \sqrt{S_x}\biggl( \frac{dx}{dT_\mathrm{load}}\biggr)^{-1}\biggl(\frac{dT_\mathrm{load}}{dP_\mathrm{rad}}\biggr)^{-1},
\end{split}
\end{equation}
where $S_x$ is determined from the flat phase noise floor at $>$20~Hz of the power spectral density as shown in Fig~\ref{fig:grid}c. The amplitude noise presented in Fig.~\ref{fig:grid}c, at a level of $-$18~dB compared to the phase noise, corresponds to the noise in the readout system \cite{Rantwijk2016}, proving that the phase noise level is not affected by the readout noise. It has to be noted that both $S_x$ and $\frac{dx}{dP_\mathrm{rad}}$ depend on background load $P_\mathrm{rad}$. We plot in Fig.~\ref{fig:grid}d the experimental NEP obtained using Eq.~\ref{eq:NEP} at $T=T_\mathrm{cold}$. In the range of 345--365~GHz where the quasioptical filter stack has its highest transmission of $\sim$40\%, the instrument has an optical NEP of $\sim$3$\times$$10^{-16}\ \mathrm{W\ Hz^{-0.5}}$.  

By equating the measured optical NEP to the theoretical optical NEP of photon-noise limited MKIDs\cite{Ferrari:bk,Flanigan:2016ef} one can obtain the total coupling efficiency between the calibration load and the detector,
\added{
\begin{equation}
\label{eq:etaopt}
	\eta_\mathrm{opt} = \frac{2P_\mathrm{rad}hF + 4\Delta P_\mathrm{rad}/\eta_{\mathrm{pb}}}{NEP^2_\mathrm{exp}-2P_\mathrm{rad}hF\tilde{n}(F,T)}
\end{equation}
}
Here, $h$ is the Planck constant, $\tilde{n}(F,T)$ is the Boltzmann occupation number of the radiation outside of the cryostat window, and $\eta_\mathrm{pb}\sim 40\%$ is the pair-breaking efficiency\cite{2014SuScT..27e5012G}. We show the optical efficiency in Fig.~\ref{fig:grid}e, together with the band-pass characteristics of the quasioptical filter stack. The figure shows that the instrument optical efficiency in the passband of the filter stack is $\eta_\mathrm{opt}\sim 2$\%.  This is close to the product of the following transmissions: 1) Quasi-optical filter stack with a transmission of $\eta_\mathrm{qof}=40\%$, 2) coupling of the cryogenic optics of  $\eta_\mathrm{co}=80\%$, simulated with physical optics using GRASP\cite{GRASP:l7_faiNy}, 3) radiation efficiency of the lens-antenna of $\eta_\mathrm{la}=70\%$, simulated in CST Microwave Studio, 4) ohmic loss of the 30 bridges across the terahertz line in between the antenna and the filterbank of $\eta_\mathrm{tl}=93\%$, estimated from an independent measurement of the loss of the bridges, and 5) on-chip filter has a mean peak transmission of $\sim$8\% as shown in Fig.~\ref{fig:toptica}c. 

It is informative to calculate the optical efficiency for the reference MKIDs, by approximating $t(F)$ with the full passband of the quasioptical filter stack normalized to unity at the maximum. The results are shown in the right panel of Fig.~\ref{fig:grid}e. 
The optical efficiency of the wideband MKIDs at the input of the filterbank is consistent with the designed value of $-27$~dB when multiplied with $\eta_\mathrm{qof}\eta_\mathrm{co}\eta_\mathrm{la}\eta_\mathrm{tl}$.
The wideband MKIDs at the output of the filterbank receive \added{$\sim$5 times} less power than those at the input, because of the fraction of power taken out by the filterbank, \added{including radiation losses, ohmic losses and reflections.} The dark MKIDs receive 30--40~dB less power than the filter channel MKIDs, showing that the amount of stray light that couples to the MKID detectors is sufficiently small for constructing a filterbank with $10^3$--$10^4$ channels.

\begin{figure*}
 \includegraphics[width=\textwidth]{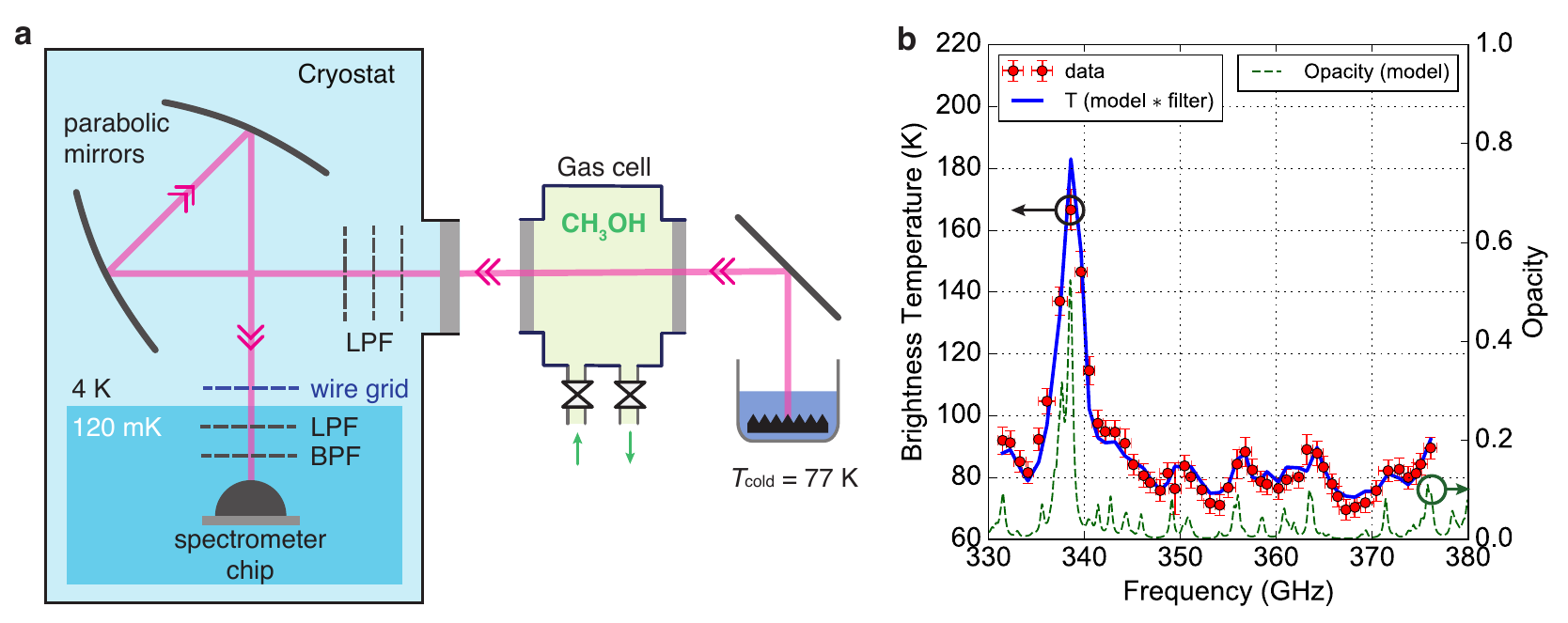}
 \caption{\label{fig:methanol} 
 \textbf{Detection of methanol gas with a single shot of a wideband terahertz spectrometer.}  
  \textbf{a}, Experimental setup. The spectrometer looks through a gas cell filled with methanol gas, into a black body load cooled with liquid $\mathrm{N}_2$.
   \textbf{b}, Wideband emission spectrum of methanol measured with the on-chip filterbank spectrometer in a single shot. The circles indicate the brightness temperature measured using the response of the spectrometer. The dashed curve is the calculated opacity \added{(defined as $1-e^{-\tau}$ where $\tau $ is the optical depth)} of methanol at a pressure of 17~mbar. The solid curve is the brightness temperature spectrum calculated by convoluting \added{this opacity} with the bandpass characteristics of the on-chip filters, with the overall amplitude and offset fitted to the measured data. The horizontal error bars represent the full width half maximum of the filter transmission of each channel. The vertical error bars represent the uncertainty in the optical spillover as estimated from the fitting in Fig.~\ref{fig:grid}b.
 }%
 \end{figure*}

\section{Detection of Methanol Gas Emission Spectrum} \label{sec:gascel}

To demonstrate that the system can spectroscopically observe emission or absorption lines from molecular gas, we have measured the emission spectrum of methanol gas at 17~mbar using the experimental configuration as shown in Fig.~\ref{fig:methanol}a. We couple the system beam to a gas cell that can be filled with methanol gas at 293 K so that the spectrometer looks through the gas cell into a black body cooled to 77 K with liquid $\mathrm{N_2}$. Initially, the gas cell is pumped to a near vacuum of 0.014~mbar. While the spectrometer is continuously observing with all channels at a sampling rate of 160~Hz, we increase the methanol pressure to 17~mbar. The relative frequency shift $\Delta x = |f-f_\mathrm{0}|/f_\mathrm{0}$, where $f_0$ is the initial MKID frequency with the gas cell at vacuum, is converted to an effective brightness temperature $T_\mathrm{b}$ by using the response curve as presented in Fig.~\ref{fig:grid}b for each MKID. To convert the individual filter channel response to a spectral brightness we use the measured spectral response of the filterbank, obtained using the photomixing CW source, as presented in Fig.\ref{fig:toptica}c. The spectral brightness obtained with our spectrometer, $T_\mathrm{b}$ is presented in Fig.~\ref{fig:methanol}b. We can compare this response to a simulation of the expected emission spectrum of methanol gas: we present the opacity of methanol gas at 17~mbar by the green dashed line in Fig.~\ref{fig:methanol}b. This spectrum is calculated using expected line frequencies and intensities from the JPL line catalog\cite{JPL_Line_Catalog} and taking different line broadening mechanisms into consideration in order to comply with the length (270 mm), pressure, and temperature of the gas cell\cite{Ren:2010ko}. By further taking into account the losses at the gas cell window, and convoluting the \added{intensity} spectrum with the bandpass characteristics of the filters as presented in Fig.~\ref{fig:toptica}c, we can calculate the expected response of each channel of the filterbank, as shown in Fig.~\ref{fig:methanol}b. Note that here we have applied to all of the spectral channels a common multiplication factor and an offset, to compensate for the uncertainty in the beam spillover at the windows of the gas cell and at the cold source. Interestingly, we find that we have to use an overall frequency shift of $-0.56$~GHz to get a maximum correspondence between the measured filterbank response and the methanol spectrum. This is consistent with the 2~GHz absolute accuracy quoted by the supplier of the CW source, and shows that it is possible to use a gas cell as a method to improve the absolute frequency calibration of the spectrometer. 

From the good correspondence between measurement and simulation as shown in Fig.~\ref{fig:methanol}b, we can conclude that the spectral shape observed by the on-chip spectrometer reproduces very well the peak frequencies and their relative strengths expected from the database.

\section{Measurement of the system beam pattern}

\begin{figure*}
 \includegraphics[width=\textwidth]{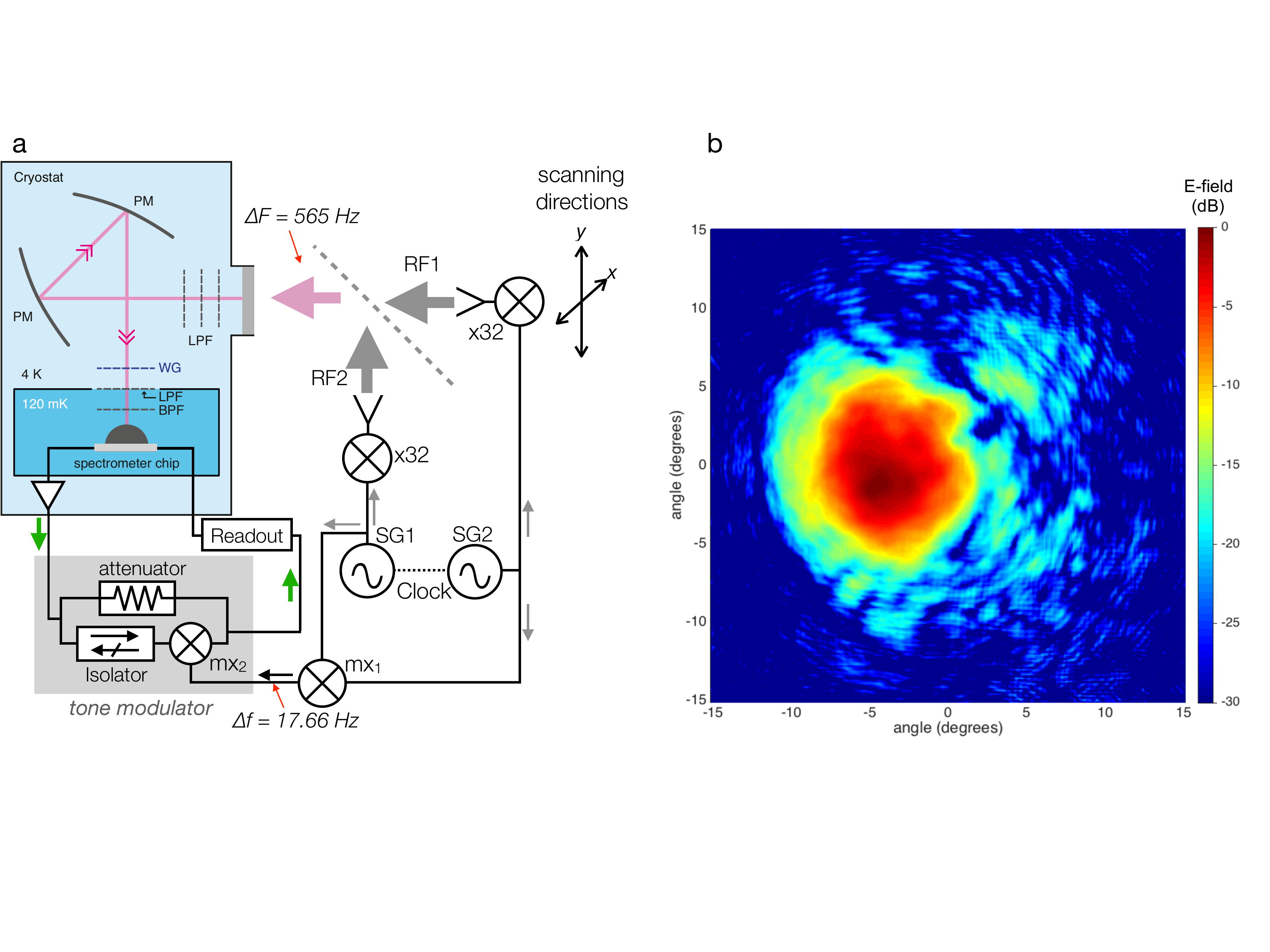}
 \caption{\label{fig:BP} 
 \textbf{The spectrometer beam pattern.}  
  \textbf{a}, Schematic of the beam pattern measurement setup, for more details see Ref.  \cite{Davis:2018IE}.  We use 2 synthesizers at a frequency of 11.25~GHz with an offset frequency $\Delta f=17.66\ \mathrm{Hz}$ to drive two $\times$32 multipliers, coupled to horns, to create 2 signals, one at RF1 at 360~GHz and RF2 at 360~GHz + 565.33~Hz. These are combined and coupled to the spectrometer using a beam splitter. Using mixer (mx1) we create a reference signal at a frequency $\Delta f$. This signal is fed to the 'tone modulator' indicated in the figure, which creates a small amplitude modulation on all readout signals coming from the spectrometer. This modulation creates a  phase reference which allows us to obtain both the phase- and amplitude signal as a function of the position of RF1.
  \textbf{b}, Far field pattern in amplitude, obtained from propagating the measured pattern at the plane of RF1 to the far field. 
 }%
 \end{figure*}

The on-chip filterbank spectrometer uses direct detectors, MKIDs, but the radiation coupling via the filterbank and lens-antenna is phase coherent. The coupling between the lens-antenna and finally the telescope is sensitive to the exact phase- and amplitude distribution of the beam, details of which are easily overlooked when doing a conventional scalar beam pattern measurement using thermal sources. To avoid this, and to be able to predict the coupling to the telescope quantitatively, we have measured the phase- and amplitude pattern, using a method pioneered in Ref. \cite{Davis:IEEE2017} and explained in Fig.~\ref{fig:BP}a. We use two coherent sources, each driven from a signal generator and a $\times$32 frequency multiplier. The two synthesizers have a small frequency difference $\Delta f=17.66\ \mathrm{Hz}$ and a source offset frequency of $\Delta F = 565\ \mathrm{Hz}$, well within the bandwidth of the detectors and readout system, which modulates the detector response in the time domain. The complex field parameters can be obtained as a function of the position of RF1 by a complex FFT of the time domain data. However, MKIDs are not phase sensitive detectors and the phase information is not conserved in the detection process. To overcome this we create a stable phase reference by adding a small amplitude modulation to all the readout tones coming from the cryostat at a frequency $\Delta f$ using the `tone modulator'. The exact technique is described in great detail in Ref.\cite{Davis:2018IE}. 

We show, in Fig.~\ref{fig:BP}b, the final result of this measurement, which is the far field pattern calculated from the measured data. We observe a reasonably clean pattern, 
which can be fitted well to a Gaussian beam, yielding a Gaussicity of 0.82, angular offsets in $x$ and $y$ given by off$_x = -4.0^\circ$  and off$_y = -0.3^\circ$,
and Gaussian beam far-field divergence angle $\theta_x = 4.4^\circ$ and $\omega_y = 5.1^\circ$, corresponding to the $1/e^2$ value in power of the best fit Gaussian beam.
The reference plane of this direction is the front side of the mounting structure of the cryostat. The cause of the relatively large angular offset is not fully understood, \added{but likely a misalignment between the field lens and the beam impinging on it.}

\section{Summary and Outlook}

We have demonstrated the operation of a full on-chip filterbank spectrometer with 49 spectral channels with a resolution of $F/\Delta F=380$ that observes over an instantaneous band of 332--377~GHz. The  photon-noise limited sensitivity of $NEP\sim3\times10^{-16}\ \mathrm{W\ Hz^{-0.5}}$, together with an actual detection of methanol gas, proves that the spectrometer can already be applied in atmospheric and astronomical sciences. Indeed, the spectrometer presented here meets all the interface requirements for immediate installation on the 10 m ASTE telescope\cite{Ezawa:2008bf} as the first generation of the DESHIMA spectrometer\cite{Endo:2012ed} for astronomy: with the excellent atmospheric transmission of the Atacama Desert at around 350~GHz, the on-chip spectrometer will more than quadruple the simultaneous bandwidth of existing heterodyne spectrometers at this type of facilities\cite{Ito:2018bb}, matching well the full spectral coverage of direct-detection cameras.

Yet the system demonstrated in this Article should be regarded as a narrow-band, single-pixel demonstrator, when compared to the vast potential scalability of the on-chip filterbank spectrometer concept. By using quarter-wavelength bandpass filters\cite{2016JLTP..184..412E} and wideband antennas\cite{2017ApPhL.110w3503B,OBrient:2013hc}, the filterbank can be naturally extended to a bandwidth of 1:3 (1.5 Octave)\cite{2012JLTP..167..341E}. 
\added{For a 1:3 broad band system the out-of-band coupling of the individual filters must be below $-40$ dB with respect to the peak power coupling to limit out-of-band power loading the detectors. The stray radiation coupling to 'blind MKIDs' in our design is already good enough, but the current filter design has spurious resonances and will not fulfill this requirement.}
Better instrument sensitivity can be reached by increasing the system coupling efficiency, by improving the coupling efficiency of the chip and the optics. The on-chip filter transmission can be improved from $\sim$8\% to values approaching unity by a combination of two methods: First, the single filter efficiency is now limited by radiation loss and ohmic losses in the aluminium bridges. \added{This can be mitigated by using microstrip filters and a microstrip THz line, which will both eliminate radiation loss, and the need for bridges.} This will bring the single filter efficiency close to the theoretical limit of 50\%\cite{Shirokoff:2012fx}. The second step requires a more advanced filter design, an example would be to combine several oversampled filters to absorb more power\cite{Shirokoff:2012fx}, and to incoherently couple that power into one MKID\cite{Bueno2017b}.
The instrument optical efficiency can be improved by using isotropic substrates (e.g., Si) for the antenna, and by a very careful choice of the infrared and quasi-optical filters.
These improvements on the chip and optics should collectively bring the instrument optical efficiency up to \added{$\eta_\mathrm{opt}\sim50$\%, corresponding to an instrument $NEP$ of $8.3\times 10^{-17}\ \mathrm{W\ Hz^{-0.5}}$.} As we show in the Appendix, an incoherent spectrometer with this NEP value will have the same per-channel sensitivity as a coherent receiver with a single-sideband receiver noise temperature of \added{$T_\mathrm{rx}\sim 22$ K:} this is very close to the standard quantum limit\cite{2011ITTST...1..241S} of $T_\mathrm{QN}=h\cdot\ 350\ \mathrm{GHz} /k=16.8\ \mathrm{K}$ for coherent detection, and a factor of $\sim$3 better than the state-of-the-art superconductor-insulator-superconductor (SIS) receivers for astronomy in this frequency range\cite{Ito:2018bb}. With the dual advantage in bandwidth and sensitivity, near-future incoherent spectrometers will offer a substantial sensitivity margin over coherent spectrometers for medium resolution spectroscopy ($F/\Delta F$$\le$ a few 1000), opening up new observational parameter space in fields such as very wide-band spectroscopy and blind spectroscopic surveys.

Furthermore, the compactness of the spectrometer chip allows small spectrometer units that can be combined into a focal plane array of spectrometer pixels---what is often referred to as a hyperspectral imager or imaging spectrometer. Indeed, the on-chip spectrometer is regarded as the most viable path towards a $\sim$100 pixel multi-object spectrometer, which is expected to substantially improve the galaxy-surveying capabilities of existing ground-based observatories for terahertz astronomy \cite{Farrah:2017wk, 2016SPIE.9906E..26K}, and enable future satellite missions for climatology and meteorology applications\cite{2017SPIE10210E..10H}.

\appendix

\added{
\section{Proof of photon noise limited radiation detection}

The noise spectrum given in Fig.~\ref{fig:grid}c is white for both phase and amplitude, which is indicative of photon noise, but not a hard proof. Referring to Ref.~\cite{Janssen:2013dg}, photon noise limited performance is proven if i) the noise is white, ii) has a roll off determined by the quasiparticle lifetime, which depends on the absorbed power. The DESHIMA system as described does not allow to perform such a test due to the limited bandwidth of the readout system. We therefore re-measured the same chip-holder combination in the exact same setup as in Ref.~\cite{Janssen:2013dg}, and measured the noise spectra using the cryogenic black body calibrator at temperature of 3~K and 30~K. A set of band-pass and low-pass filters defining a 320--380~GHz band coupling between calibration load and the DESHIMA holder, the power to an individual MKID, taking into account the DESHIMA filter bandwidth, is 0.015~fW at 3~K and 15~fW at 30~K, the power absorbed in DESHIMA is 20~fW at a 77~K load. The resulted spectrum from one of the resonators is shown in Fig.~\ref{fig:ADR_PSD}. We clearly observe a white noise spectrum, with a roll-off depending on the absorbed power. At the 3~K black body temperature, the noise is still not quite flat, indicating that two-level-system (TLS) noise still has a small contribution. At 30~K, the noise is white in the 10--100~Hz frequency range used in the main manuscript to determine he coupling efficiency. Also, we clearly see a power dependent roll off in the spectrum. This shows that the device is intrinsically photon noise limited at all power levels relevant for the experiment. Note that the amplitude noise shows a similar behavior, but the photon signature exceeds the amplifier noise, given by the level at the highest frequency, only marginally, implying a significant system noise contribution even at low frequencies.

\begin{figure*}
\centering
 \includegraphics[width=0.6\textwidth]{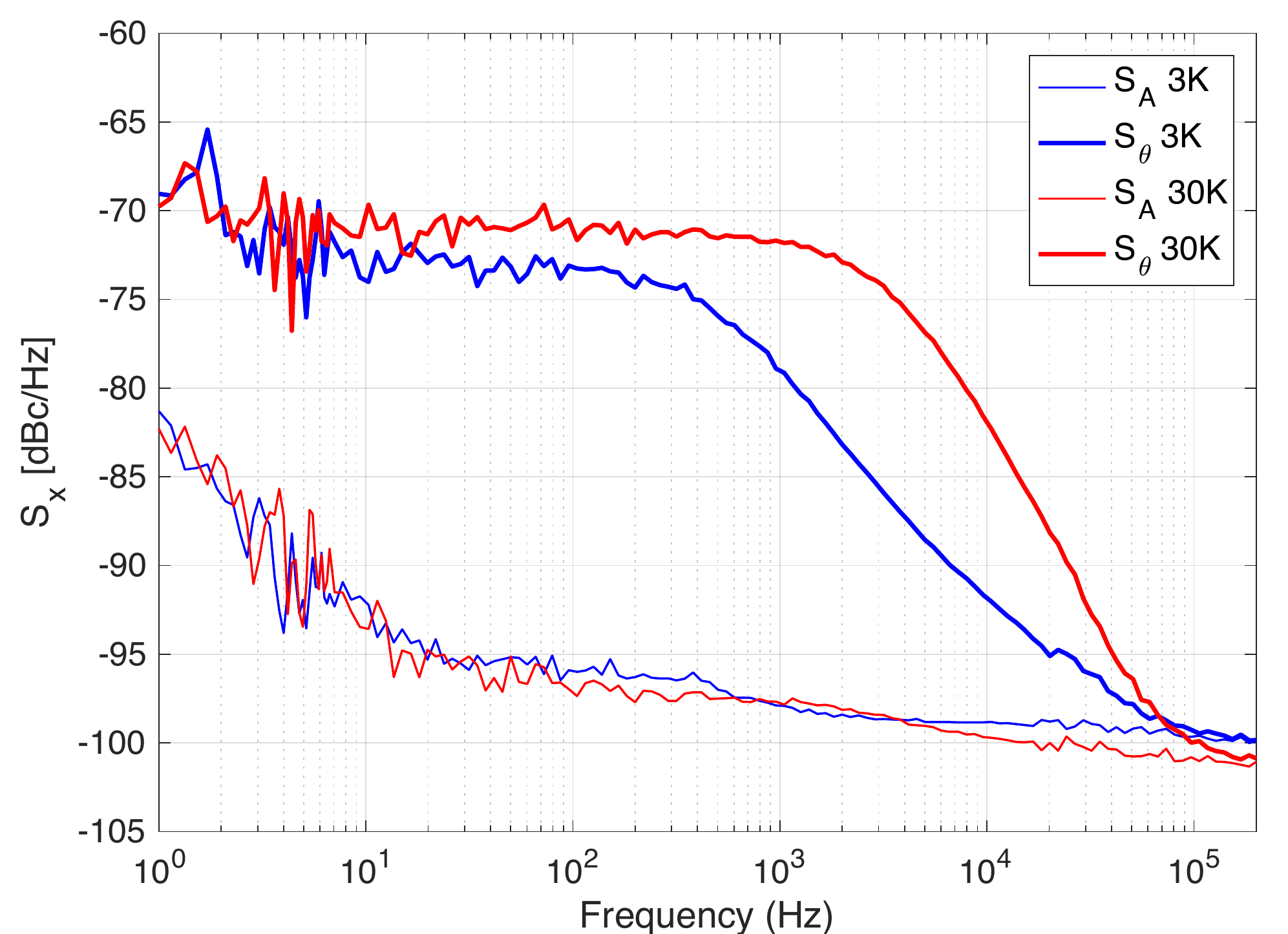}
 \caption{\label{fig:ADR_PSD} 
 \textbf{Photon noise observed by an MKID in the filterbank.}  
     \added{The power spectral density of the amplitude (dashed line) and phase (solid line) noise measured under thermal optical loading of 3 K (blue) and 30 K (red). A white noise spectrum is observed for phase readout, the level of which is constant with loading power, for $T_\mathrm{load}>3$ K. The roll-off above 1 kHz is due to the quasiparticle lifetime.}
 }%
 \end{figure*}

}

\section{Equivalent coherent receiver noise temperature of a MKID-based incoherent spectrometer}

The calculation is similar to the case in which the incoherent system is completely photon-noise limited\cite{2011ITTST...1..241S}, except that here we include the relatively small effect of the quasiparticle recombination noise of MKIDs. We will consider a case representative for a spectrometer system installed on a ground-based astronomical observatory, where the signal to  be detected has a power that is much smaller compared to a background temperature of $T_\mathrm{bkg}=77$ K typical for the 0.3--1.0 THz range\cite{2011ITTST...1..241S}. 

The signal to noise ratio (SNR) of a coherent receiver is given by the radiometer equation:
\begin{equation}
\begin{split}
SNR_\mathrm{coh}&= \frac{T_\mathrm{S}}{T_\mathrm{N}}\sqrt{\tau\cdot \Delta F},\\
\end{split}
\end{equation}
where $T_\mathrm{S}$ is the brightness temperature of the signal, $T_\mathrm{N}$ is the system noise temperature, $\tau$ is the integration time, and $\Delta F$ is the detection bandwidth. $T_\mathrm{N}$ is the sum of $T_\mathrm{bkg}$ and the receiver noise temperature $T_\mathrm{rx}$. 

Similarly, the SNR for an incoherent spectrometer is given by:
\begin{equation}
\begin{split}
SNR_\mathrm{inc}&= \frac{P_\mathrm{S}}{NEP}\sqrt{2\tau},\\
\end{split}
\end{equation}
where $P_\mathrm{S}$ is the power of the signal. 

Equating $SNR_\mathrm{coh}=SNR_\mathrm{inc}$ using $P_\mathrm{S}\sim k  \cdot \Delta F \cdot T_\mathrm{S}$ and $T_\mathrm{N}=T_\mathrm{bkg}+T_\mathrm{rx}$, we obtain 

\begin{align}
\label{eq:Trx}
T_\mathrm{RX} &= \frac{NEP}{k}\frac{1}{\sqrt{2\Delta F}}-T_\mathrm{bkg}.
\end{align}
Note that $T_\mathrm{RX}$ is independent of $\Delta F$ for a photon-noise limited MKID, because the NEP scales with $\sqrt{\Delta F}$. 

As discussed in the main text, an on-chip filterbank spectrometer with an instrument optical efficiency of \added{$\eta_\mathrm{opt}=50\%$ yields a NEP of $8.3\times 10^{-17}\ \mathrm{W\ Hz^{-0.5}}$.} Here we have taken $\Delta F=0.5\pi\cdot 350\ \mathrm{GHz}/Q$ and $Q=300$, where the factor $0.5\pi$ is the area under a Lorentzian curve whose peak amplitude and full-width-half-maximum are both unity. According to Eq.~\ref{eq:Trx}, this NEP would correspond to a $T_\mathrm{rx}$ of \added{22~K}, which is close to the standard quantum limit for coherent detection\cite{2011ITTST...1..241S}.

\subsection*{Disclosures}
The authors have no relevant financial interests in the manuscript and no other potential conflicts of interest to disclose.

\acknowledgments 
We thank Peter Hargrave for suggesting promising applications of the on-chip spectrometer to atmospheric sciences, and Klaas Keizer for the precise mechanical work on the cryostat. This research was supported by the Netherlands Organization for Scientific Research NWO (Vidi grant No. 639.042.423, NWO Medium Investment grant No. 614.061.611 DESHIMA), the European Research Counsel ERC (Consolidator grant No. 648135 MOSAIC), and the Japan Society for the Promotion of Science JSPS (KAKENHI Grant Numbers JP25247019 and JP17H06130). P.J. de V. is supported by the NWO (Veni Grant 639.041.750). T.M.K. is supported by the ERC Advanced Grant No. 339306 (METIQUM) and the Russian Science Foundation (Grant No. 17-72-30036). N.L. is supported by ERC (Starting Grant No. 639749). J.S. and M.N. were supported by the JSPS Program for Advancing Strategic International Networks to Accelerate the Circulation of Talented Researchers (Program No. R2804). B.M. was supported by the European Union’s Horizon 2020 research and innovation program under grant agreement No 730562 (RadioNet).


\bibliography{ms}   
\bibliographystyle{spiejour}   


\vspace{2ex}\noindent\textbf{Akira Endo} is an assistant professor in the Terahertz Sensing Group of TU Delft. He is interested in studying galaxies at high redshift using superconducting detectors. He obtained his PhD (astronomy) from the University of Tokyo in 2009.  He has been focusing on the DESHIMA experiment since 2009, when he began a postdoc at the Kavli Institute of Nanoscience Delft. In his current position (since 2014), he also enjoys teaching astronomical instrumentation and systems engineering.

\vspace{2ex}\noindent\textbf{Kenichi Karatsu} is a post-doctoral researcher at SRON.
He received his Ph.D. degree in science in 2011 from Kyoto University with study of proton spin structure at the RHIC-PHENIX experiment.
In 2015, He joined the DESHIMA project at TU Delft/SRON.
His main role is to lead laboratory evaluation and telescope campaign of the instrument.
His research interest is to develop an experimental instrument for revealing mysteries of the Universe. 

\vspace{2ex}\noindent\textbf{Alejandro Pascual Laguna} received the B.Sc. degree in Telecommunications Engineering from Universidad Pontificia Comillas, ICAI School of Engineering, Madrid, Spain, in 2014. In 2016 he received the M.Sc. degree (cum laude) in Electrical Engineering from Delft University of Technology, Delft, The Netherlands, where he is currently working toward the Ph.D. degree with the Terahertz Sensing group.
His research interests include on-chip solutions for efficient broadband sub-mm wave spectrometric and imaging systems based on Kinetic Inductance Detectors.

\vspace{2ex}\noindent\textbf{Behnam Mirzaei} obtained his PhD from the Quantum Nanoscience department of Delft University of Technology (TU Delft) in 2018. His focus was on the application of quantum cascade lasers (QCLs) as local oscillators (LO) in super-THz heterodyne receivers. He is currently contributing to the development of the receiver unit of GUSTO (Galactic/extragalactic Ultra long duration balloon Spectroscopic Stratospheric THz Observatory), as a post-doc in TU Delft together with a team in SRON (Netherlands Institute for Space Research)

\vspace{2ex}\noindent\textbf{Robert Huiting} is a mechanical design engineer. He received his BS degree precision engineering (2001) and technology management (2003) at the Hogeschool van Utrecht. He has been working as a design engineer for SRON Netherlands institute for space research since 2007 He has worked as a research engineer at the FOM institute for plasma physics for the XUV optics group the years before. His experience from the last 6 years lies in designing instrumentation for ground based space research.

\vspace{2ex}\noindent\textbf{David J. Thoen} is a cleanroom engineer  working at Delft University of Technology (TUD). He received his BS (2008)  degree in Applied Physics at Fontys University of Applied Sciences, Eindhoven  while he worked as a microwave engineer at FOM Institute Rijnhuizen since 2007. Since 2010 he is working at TUD on the development of Microwave-kinetic-inductance-detectors (MKID) far-infrared detectors for astronomy. He has extensive experience in cleanroom processing, process development, vacuum and cryogenic technology.

\vspace{2ex}\noindent\textbf{Vignesh Murugesan} received his M.Sc. in Microsystem integration technology from Chalmers University of Technology (2007). He worked as a process integration engineer from 2007 to 2008 for Infineon Technologies, Germany. From 2010 to 2013 he worked as a MEMS process engineer for Thermo Fisher Scientific, Netherlands. Since 2013, he has been working as a process engineer at SRON Netherlands for MKIDS group. He is currently responsible for the fabrication and process development of MKID chips.

\vspace{2ex}\noindent\textbf{Stephen Yates} is an instrument scientist at SRON, the Netherlands
Institute for Space Research (since 2006). He received a Ph.D. from the
University of Bristol  (2003), followed by work at the CNRS-CRTBT (now
Institut Ne\'el) Grenoble on experimental low temperature techniques
(2003-2006). His current interests concentrate on MKID development for
sub-mm astronomy applications but also include a wider interest in
device physics and superconductivity, optics, and full end to end
instrument characterization and performance.

\vspace{2ex}\noindent\textbf{Juan Bueno} graduated in Physics from the University of Cantabria in 2003 and received his Ph.D. degree at the University of Leiden in 2007. In 2008, he was awarded with a NASA Postdoctoral position, joining the Jet Propulsion Laboratory, where he pioneered a new type of pair-breaking radiation detector, the Quantum Capacitance Detector. He became an instrument scientist at SRON in 2012 working on the development of far-IR and sub-mm wave Kinetic Inductance Detectors.

\vspace{2ex}\noindent\textbf{Nuri van Marrewijk} received his MSc. degree in Applied Physics from the Kavli Institute of Nanoscience, TU Delft, in 2014. He participated in the DESHIMA project as PhD student at the Faculty of Electrical Engineering, Mathematics and Computer Science, TU Delft.

\vspace{2ex}\noindent\textbf{Sjoerd Bosma} received his B.Sc. (2015) and M.Sc. (2017, cum laude) degrees in electrical engineering from the Delft University of Technology, Delft, The Netherlands (TU Delft). He is currently a Ph.D. candidate at the Terahertz Sensing Group at the TU Delft where he works on leaky-wave lens antenna arrays for submillimeter-wave spectrometers. From September 2018 to February 2019 he participated in the JPL Visiting Student Researcher Program at the Jet Propulsion Laboratory, California, USA.

\vspace{2ex}\noindent\textbf{Ozan Yurduseven} is a senior antenna engineer at Huawei Technologies Dusseldorf GmbH, Germany. He received his B.Sc. and M.Sc. (Hons.) degrees in Electronics and Communications Engineering from Yildiz Technical University, Istanbul, Turkey, in 2009 and 2011, respectively. He obtained his Ph.D. degree in Microelectronics Department from the Delft University of Technology, Delft, The Netherlands, in 2016. His research interests include wideband integrated lens antennas for THz imaging, numerical techniques in electromagnetics and metamaterials.

\vspace{2ex}\noindent\textbf{Nuria Llombart} is a professor at the THz Sensing Group of TU Delft, where she has been working since 2012. Her current research interests include the analysis and design of planar antennas, periodic structures, reflector antennas, lens antennas, and waveguide structures, with emphasis in the terahertz range. In 2015, she was the recipient of European Research Council Starting Grant. She serves as a Board Member of the IRMMW-THz International Society.

\vspace{2ex}\noindent\textbf{Junya Suzuki} is a postdoctoral fellow in the Institute of Particle and Nuclear Studies (IPNS) at KEK. He received his BS (2011) degree in physics from the University of Tokyo, and his MS (2013) and Ph.D. (2016) degrees in Physics from the University of Tokyo. He joined IPNS in April 2016 and is involved in the development of mm-wave telescopes.

\vspace{2ex}\noindent\textbf{Masato Naruse} is an assistant professor in the Graduate School of Science and Engineering at Saitama University. He received his BS (2007) degree in physics from Kyoto University, and his MS (2009) and his PhD (2012) degrees in science from the University of Tokyo. He joined the Saitama University faculty in 2012 and develops high sensitivity superconducting detectors for astronomy and industry.

\vspace{2ex}\noindent\textbf{Pieter J. de Visser} is an instrument scientist at SRON, Netherlands Institute for Space Research. He received his BS (2007) and MS (2009) degrees in applied physics from Delft University of Technology with a specialization in astronomy \& instrumentation at Leiden University. He received his PhD degree (2014) from Delft University of Technology (cum laude). After a Postdoc at the University of Geneva, Switzerland, he now focuses at SRON on energy-resolving detectors for visible/near-infrared light.

\vspace{2ex}\noindent\textbf{Paul P van der Werf} is Professor of Extragalactic Astrophysics at Leiden Observatory, Leiden University, The Netherlands. His interests are in the
interstellar medium of galaxies at low and high redshift, in the cosmic evolution of the galaxy population, and in the fundamentals of science. His observational
work focuses on spectroscopy in wavelength regions from the near-infrared to the radio regime. 

\vspace{2ex}\noindent\textbf{Teunis M. Klapwijk} is at the Kavli Institute of Nanoscience, mostly known for his fundamental contributions to superconductivity. He obtained his PhD degree in 1977, continued his work at Delft until 1985, with a break at Harvard in 1979-1980, followed by a professorship at Groningen.  He returned to Delft in 1999, where he became a research-only professor in 2013. He has also been a key-contributor to research on superconductors for astronomical detection.      

\vspace{2ex}\noindent\textbf{Jochem Baselmans} (1974) is senior instrument scientist at the SRON Netherlands Institute for Space Research and professor at Delft University of Technology. He received his PH.D. (summa cum laude) at the university of Groningen on 2002, entitled: ‘Controllable Josephson Junctions’. He leads the Dutch effort on the development of Microwave Kinetic Inductance Detectors, where his main interests are ultra-sensitive MKIDs for THz radiation detection and advanced on-chip imaging spectrometers for sub-THz imaging spectroscopy

%


\end{spacing}
\end{document}